\documentclass[12pt]{article}
\usepackage{graphicx}
\usepackage{amsmath}
\usepackage{amsbsy}
\usepackage{amsfonts}
\usepackage{amsthm}
\begin{document}

\title{Domain walls on the surface of q-stars}

\author{Athanasios Prikas}

\date{}

\maketitle

Physics Department, National Technical University, Zografou
Campus, 157 80 Athens, Greece.\footnote{e-mail:
aprikas@central.ntua.gr}

\begin{abstract}
We study domain-wall networks on the surface of q-stars in
asymptotically flat or anti de Sitter spacetime. We provide
numerical solutions for the whole phase space of the stable field
configurations and find that the mass, radius and particle number
of the star is larger but the scalar field, responsible for the
formation of the soliton, acquires smaller values when a
domain-wall network is entrapped on the star surface.
\end{abstract}

PACS number(s): 11.27.+d, 04.40.-b, 11.30.Er, 97.10.Cv

\newpage

\section{Introduction}

Planar domain wall networks in theories with three or more vacua
have been investigated in a series of papers,
\cite{dw1,dw2,dw3,dw4}. Domain walls seem to have various
applications especially in D-brane theories,
\cite{dw-brane1,dw-brane2}.

The idea of 3-dimensional networks on spherical surfaces has been
established in \cite{dw-spherical} and applied to the spherical
surface of a ``large" soliton star, \cite{dw-stars}. ``Large"
soliton stars were investigated in a series of papers by Friedberg
\textit{et al},
\cite{large-soliton-stars1,large-soliton-stars2,large-soliton-stars3,large-soliton-stars4}.
In \cite{dw-stars} it was found that there exist stable with respect
to fission into free particles field configurations, corresponding
to soliton stars, with a domain wall network on their surface. The
pressure of the soliton acts as a stabilizing force to the domain
wall network. The star mass is, in general, slightly larger in the
presence of the above network, when the particle number and the
radius show a more complicated behavior, depending on the surface
tension, the particle number etc. In \cite{dw-solitons} it was
discussed the entrapment of domain wall network on the surface of a
soliton of any kind in the absence of gravity. In the same article,
another kind of ``stability" was discussed, namely the stability of
the domain walls. It was found that only two of the five Platonic
solids, cube and octahedron, can be realized on a spherical surface.
Networks forming other solids would collapse to a single vacuum. So,
from now on, when discussing domain walls we restrict ourselves to
these two solids.

In the present article we investigate the properties of q-stars
\cite{qstars-global,qstars-charged,qstars-extradim}, which are
relativistic generalizations of q-balls, \cite{qballs-initial}, that
seem to play a special role in baryogenesis through flat directions
of supersymmetric extensions of Standard Model,
\cite{qballs-baryogen}. Our purpose is to find numerical solutions
to the coupled Einstein-scalar fields equations, and to calculate
the mass, radius, particle number, and the value of the scalar field
at the center of the soliton in the whole phase space. We compare
our results with those obtained in the absence of the domain wall
network, including in our figures the relevant results, and to the
results of \cite{dw-stars}, referring to the other family of
solitonic stars, and \cite{dw-solitons}. We also investigate the
soliton stability with respect to gravitational collapse and to
decay to free particles.

\section{Domain wall network on the surface of q-stars}

We consider a complex scalar field, $\phi$, with a global $U(1)$
symmetry and a suitable potential $U$, coupled to a complex scalar
field, $\psi$, able to generate a network of domain walls, and to
gravity. In order the $\psi$ field to be able to generate a
domain-wall network, a $$\left|a-\frac{\psi^n}{v^n}\right|^2$$
term should be included in the action, where $a$ is a constant,
\cite{dw1,dw2,dw3,dw4,dw-brane1,dw-brane2}. The action of the
above configuration is:
\begin{align}\label{1}
S=\int d^4x\sqrt{-g}\left[\frac{R-2\Lambda}{16\pi
G}+\partial_{\mu}\phi\partial^{\mu}\phi^{\ast}+
\frac{1}{v^2}\partial_{\mu}\psi\partial^{\mu}\psi^{\ast} \right. \nonumber\\
\left.
-U(\phi^{\ast}\phi)-\beta^2\left|c(f_0-\phi^{\ast}\phi)\phi^{\ast}\phi
-\frac{\psi^n}{v^n}\right|^2 \right]
\end{align}
where $\Lambda$ stands for the (negative or zero) cosmological
constant and the constants $\beta$, $c$ and $f_0$ will be
determined later, $n$ is an integer and $v$ is a positive constant
that can be absorbed in $\psi$ and does not affect the solutions.
A Lagrangian with a potential of the form:
$$\left|1-\frac{\psi^n}{v^n}\right|^2$$ admits solutions
corresponding to a domain wall network. In our case we choose:
\begin{equation}\label{2}
\phi(\vec{\rho},t)=\sigma(\rho)e^{-\imath\omega t}\ .
\end{equation}
The field configuration is spherically symmetric, so we can choose
a spherically symmetric metric:
\begin{equation}\label{4}
ds^2=-\frac{1}{B(\rho)}dt^2+\frac{1}{A(\rho)}
d\rho^2+\rho^2d\vartheta^2+\rho^2\sin^2\vartheta d\varphi^2\ .
\end{equation}
We define:
\begin{equation}\label{5}
\begin{split}
W\equiv B{\left(\frac{\partial\phi}{\partial
t}\right)}^{\ast}\left(\frac{\partial\phi}{\partial t}\right)=
B{\omega}^2{\sigma}^2\ ,\hspace{1em}  V\equiv
A{\left(\frac{\partial\phi}{\partial\rho}\right)}^{\ast}
\left(\frac{\partial\phi}{\partial\rho}\right)= A{\sigma'}^2
\end{split}
\end{equation}
and make the following rescalings:
\begin{equation}\label{6}
\begin{split}
&\tilde{\rho}=\rho m\ , \hspace{1em} \tilde{\omega}=\omega/m\ ,
\hspace{1em} \tilde{\phi}=\phi/m \
,\hspace{1em}\tilde{r}=\epsilon\tilde{\rho}\ , \\
&\tilde{\beta}=\beta/m^2\ ,\hspace{1em} \tilde{f_0}=f_0/m^2\
,\hspace{1em}
\tilde{c}=cm^4 \\
&\widetilde{U}=U/m^4\ , \hspace{1em} \widetilde{W}=W/m^4\ ,
\hspace{1em} \widetilde{V}=V/m^4\
,\hspace{1em}\widetilde{\Lambda}\equiv \frac{\Lambda}{8\pi Gm^4}\
,
\end{split}
\end{equation}
with $$\epsilon\equiv\sqrt{8\pi Gm^2}\ ,$$ a very small quantity
for $m\sim GeV$. Quantities of the same order of magnitude as
$\epsilon$ can be neglected. We also choose a rescaled potential:
\begin{equation}\label{3}
U=\phi^{\ast}\phi\left(1-\phi^{\ast}\phi+\frac{1}{3}(\phi^{\ast}\phi)^2\right)=\sigma^2
\left(1-\sigma^2+\frac{1}{3}\sigma^4\right)\ ,
\end{equation}
where we dropped tildes for simplicity.

We will now determine the proper values of $f_0$ and $c$ regarding
at present $\beta=0$. Gravity becomes important when $R\sim GM$.
If we regard that the scalar field $\sigma$ varies very slowly
with respect to the radial coordinate then
$d\sigma/d\rho\sim\epsilon$ if we set $m=1$. So, the
Euler-Lagrange equation for the scalar field, dropping the tildes
and the $O(\epsilon)$ quantities gives:
\begin{equation}\label{7}
\sigma^2=1+\omega B^{1/2}\ .
\end{equation}
The eigenvalue equation for the frequency can be obtained by
integrating the equation of motion within the surface, where the
scalar field varies rapidly from a $\sigma_0$ value at the inner
edge of the surface, to a zero value at the outer edge. The result
of integration is:
\begin{equation}\label{8}
V+W-U=0\ ,
\end{equation}
which, in order to match with the interior solution, for which
$\sigma'\sim\epsilon$, gives for the eigen-frequency:
\begin{equation}\label{9}
\omega=\frac{A_{\textrm{sur}}^{1/2}}{2}=\frac{B_{\textrm{sur}}^{-1/2}}{2}\
,
\end{equation}
where $A_{\textrm{sur}}$, $B_{\textrm{sur}}$ denote the value of
the metrics at the surface of the star. In the absence of gravity
$B(r)=1$ and, as one can find from eqs. \ref{7} and \ref{9},
$\sigma^2=1.5$. When gravity is under consideration the situation
is not too different, unless $B(r)\gg1$. When
$B(0)\rightarrow\infty$ the star collapses to a black hole.
Excluding this case and regarding only stars in our discussion, we
can set $f_0=1.5$, which means that, the quantity
$c(f_0-\phi^{\ast}\phi)\phi^{\ast}\phi$ is zero inside and outside
the soliton. The only region within which this quantity differs
form zero is the very thin surface. The maximum value is at
$$\sigma^2=\frac{3}{4}\ .$$ So we set:
$$c=\frac{16}{9}$$ and the above quantity takes its proper maximum value:
$c(f_0-\phi^{\ast}\phi)\phi^{\ast}\phi|_{\max}=1$. With these
settings the field $\psi$ is exactly zero outside the soliton,
approximately zero in the interior and maximum within the surface.

\begin{figure}
\centering
\includegraphics{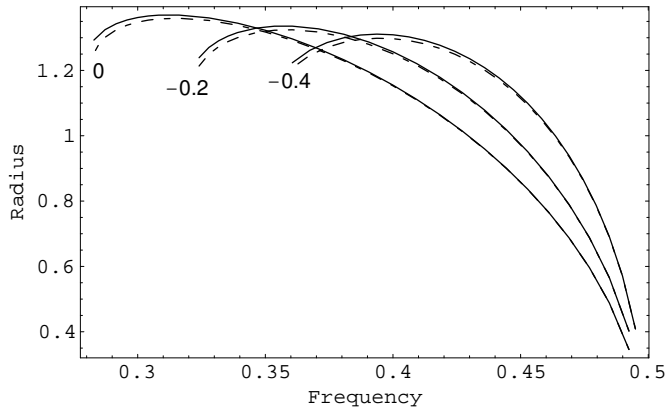}
\caption{The radius of the q-star as a function of the frequency.
Solid lines correspond to the case with domain walls and dashed to
the case without domain walls, i.e. $\beta=0$, included for
comparison. The numbers within the figures denote the value of the
cosmological constant $\Lambda$. The true cosmological constant is
$\Lambda\times8\pi Gm^4$, according to eq. \ref{6}.}
\label{figure1}
\end{figure}

\begin{figure}
\centering
\includegraphics{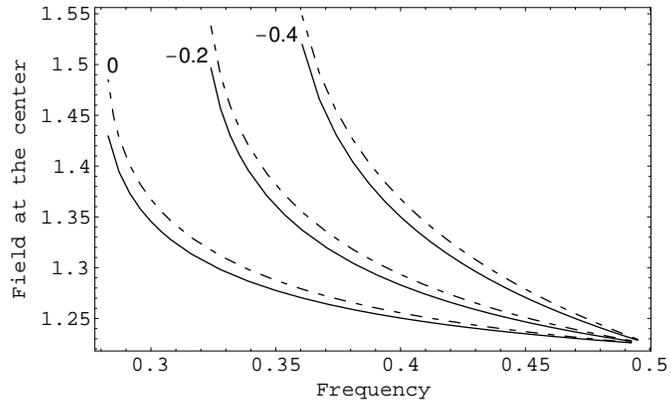}
\caption{The value of the scalar field at the center of the star
as a function of its frequency.} \label{figure2}
\end{figure}

\begin{figure}
\centering
\includegraphics{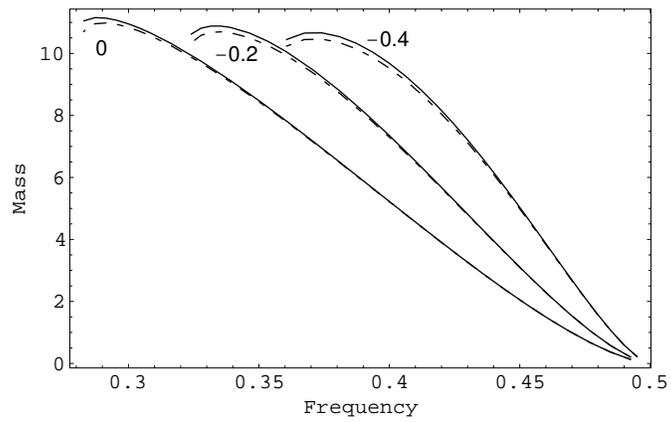}
\caption{The mass of a q-star as a function of its frequency.}
\label{figure3}
\end{figure}

\begin{figure}
\centering
\includegraphics{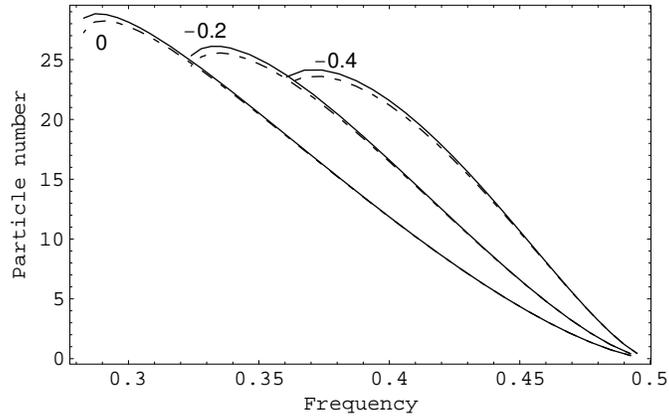}
\caption{The particle number of a q-star as a function of its
frequency.} \label{figure4}
\end{figure}

\begin{figure}
\centering
\includegraphics{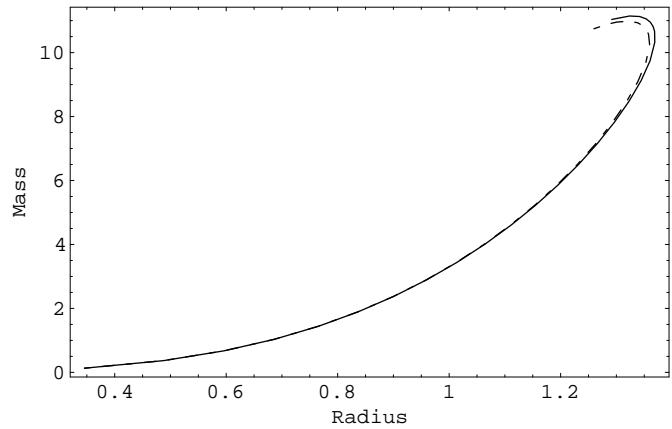}
\caption{The mass of the star as a function of its radius for
asymptotically flat spacetime. The last stable field
configurations with respect to gravitational collapse are at the
top of the $M=M(R)$ curves.} \label{figure5}
\end{figure}

\begin{figure}
\centering
\includegraphics{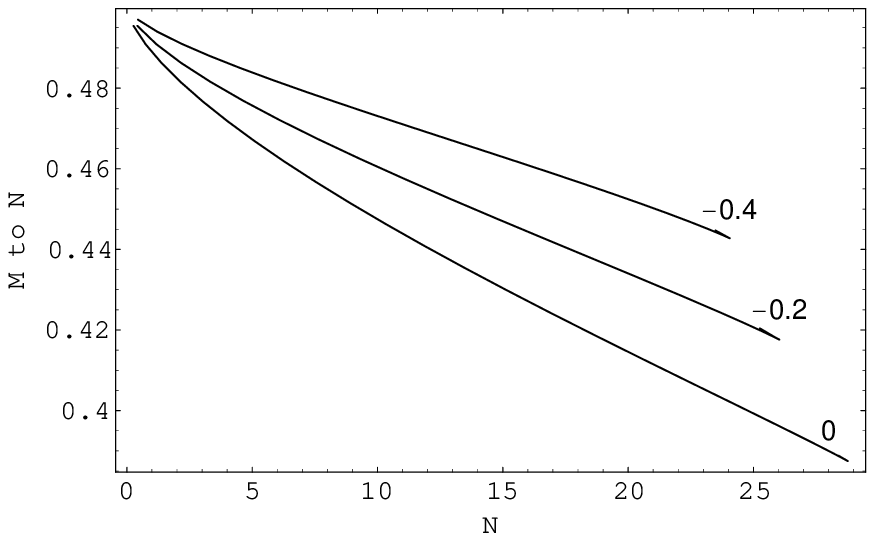}
\caption{The ratio mass to particle number of the star as a function
of the particle number of the field configuration, for
asymptotically flat or anti de Sitter spacetime. For any curve,
$\beta^2=0.01$ holds.} \label{figure6}
\end{figure}

We will now turn on the interactions between the two scalar
fields. Let $\beta^2=0.01$ so as to treat the additional
interaction as a perturbation, which does not disturb
significantly the system. The equation of motion for the $\phi$
field, dropping the $O(\epsilon)$ quantities gives:
\begin{equation}\label{10}
\omega^2B-(\sigma^2-1)^2\left(1+\frac{128}{9}\beta^2\sigma^2\right)+\frac{128}{81}
\beta^2\sigma^2=0\ .
\end{equation}
The exact solution in the above equation is rather long and ugly,
but if we substitute the value $\sigma^2=1.5$ (holding true when
$B(r)\sim1$ and $\beta^2\ll$) in the $\beta^2\sigma^2$
combination, we find for $\sigma^2$:
\begin{equation}\label{11}
\sigma^2=1+\sqrt{\frac{\omega^2B+16\beta^2/3}{1+64\beta^2/3}}\ ,
\end{equation}
which has the right limiting value for $\beta\rightarrow0$
according to the eq. \ref{7}. Integrating again the equation of
motion within the surface, we find that the eigenvalue equation
for the frequency, eq. \ref{9}, remains valid.

Within the soliton, the $\psi$ field is approximately zero. So,
the total energy-momentum tensor is approximately the
energy-momentum tensor for the $\phi$ field, which takes the form:
\begin{equation}\label{12}
T_{\mu\nu}={({\partial}_{\mu}\phi)}^{\ast}({\partial}_{\nu}\phi)+
({\partial}_{\mu}\phi){({\partial}_{\nu}\phi)}^{\ast}
-g_{\mu\nu}[g^{\alpha\beta}{({\partial}_{\alpha}\phi)}^{\ast}({\partial}_{\beta}\phi)]
-g_{\mu\nu}U\ .
\end{equation}

The Einstein equations are $G_{\mu\nu}=8\pi GT_{\mu\nu}-\Lambda
g_{\mu\nu}$. The independent components, $00$ and $11$, take the
following form, dropping the tildes and the $O(\epsilon)$
quantities:
\begin{equation}\label{13}
\frac{A-1}{r^2}+\frac{A'}{r}=-W-U-\Lambda\ ,
\end{equation}
\begin{equation}\label{14}
\frac{A-1}{r^2}-\frac{B'}{r}\frac{A}{B}=W-U-\Lambda\ ,
\end{equation}
where $U=\sigma^2-\sigma^4+1/3\sigma^6$ and $W=B\omega^2\sigma^2$
and $\sigma$ is given by eq. \ref{11}. The mass of the field
configuration is given by the relation:
\begin{equation}\label{15}
M=4\pi r\left(1-A(r)-\frac{1}{3}\Lambda r^3\right)\
,\hspace{1em}r\rightarrow\infty\ .
\end{equation}
The particle number is defined as:
\begin{equation}\label{16}
N\equiv\int d^3xj^0\ ,
\end{equation}
with:
\begin{equation}\label{17}
j^{\mu}=\sqrt{-g}g^{\mu\nu}\imath(\phi^{\ast}\partial_{\nu}\phi-\phi\partial_{\nu}\phi^{\ast})\
.
\end{equation}
For the interior, we take for the particle number:
\begin{equation}\label{18}
N=8\pi\int_0^R r^2dr\omega\sigma^2\sqrt{\frac{B}{A}}\ ,
\end{equation}
where $R$ is the radius of the q-star.

We will now prove that the energy and particle number
contributions form the thin surface are negligible. At the
exterior, the $\phi$ field is exactly zero, and consequently the
$\psi$ field is zero, and, thus, no energy or particle number
contribution arises from the exterior. The surface is of width of
order $m^{-1}$. For the moment we ignore our rescalings. Within
the surface $|\phi|\sim m$ and consequently $U\sim W\sim m^4$.
Also, the field $\phi$ varies from a certain value at the inner
edge of the surface to a zero value at the outer, so $V\sim
(|\phi| m)^2\sim m^4$. $\psi^n/v^n$ is of the same order of
magnitude as $c|\phi^4|$. Because $\tilde{c}\sim1$, then $c\sim
m^{-4}$. Because $\tilde{\beta}^2\sim10^{-2}$, then $\beta^2\sim
m^410^{-2}$, which means that within the surface
$\psi^n/v^n\sim10^{-2}m^4$. The $\psi$ field contributes to the
total energy through the $\beta^2$ terms of potential energy and
the kinetic terms. The energy density arising from the $\beta^2$
terms are of the same order of magnitude as $\beta^2m^4$. If we
repeat the discussion concerning the kinetic terms for the $\phi$
field, the kinetic terms for the $\psi$ field are of the same
order of magnitude as $\beta^2m^4$. So, the total energy density
from the surface (potential for both fields, kinetic from the
temporal variation from the field $\phi$ and kinetic from the
spatial variation from both fields) is $\sim m^4$ and the total
energy stored within the surface is $\sim1/(Gm)$. Within the
interior the energy density is $\sim m^4$ and the total energy is
$\sim1/(G^{3/2}m^2)$. One can see that
$E_{\textrm{surface}}\sim\epsilon E_{\textrm{interior}}$ and thus
the contribution of the surface to the total energy is negligible.
The same discussion holds true for the particle number
contribution of the surface. These steps reproduce the similar
discussion holding true in the case of a ``pure" q-star without
any additional fields in the total action.

We numerically solve equations \ref{13} and \ref{14} with boundary
conditions $A(0)=1$, $A(r)=1/B(r)=1-(1/3)\Lambda r^2\
,r\rightarrow\infty$ and find the parameters of the q-star using
relations \ref{15}-\ref{18} and \ref{11}.

\section{Concluding remarks}

In the present work we investigated the formation of q-stars with a
domain wall network on their surface. All the field configurations
are stable with respect to fission into free particles because the
energy of the free particles with the same charge is smaller than
the star mass. They are also stable with respect to gravitational
collapse, as one can see from figure \ref{figure5}.

We find that the radius, mass and particle number of the q-star
are slightly larger when domain walls are trapped on the star
surface. These results agrees with the estimation of
\cite{dw-stars}, concerning the total energy of the soliton.
Radius and particle number of ``large" soliton stars, investigated
in \cite{dw-stars} show a more complicated dependence on
$\beta^2$, which is the new parameter differentiating the stars
with domain walls from usual, ``large" soliton stars. The results
of larger radii and smaller values of the scalar field at the
center of the star agree with the similar ones, obtained in
\cite{dw-solitons}, despite the absence of gravity. The same
results hold true in asymptotically flat, as well as in
asymptotically anti de Sitter spacetime.

Unfortunately, one can not find an analytical relation connecting
the mass and the particle number of the star, due to the highly
non-linear character of the equations of motion. Instead, we use our
numerical results, depicted in figure \ref{figure6}, in order to
study the behavior of the soliton mass as a function of its particle
number. From figure \ref{figure6} one can see that our solutions are
always stable with respect to fission into free particles, because
the soliton energy is always smaller than the total energy of the
free particles. For small values of the particle number, gravity is
negligible. In the absence of gravity and for a large soliton the so
called thin-wall approximation holds. It has been shown that for
such solitons $M=\omega N$ plus some surface terms which are
negligible in the above approximation, with
$\omega\equiv\left(\sqrt{U/\sigma^2}\right)_{\textrm{min}}$. As one
can see, for the potential \ref{6}, $\omega$ equals to 0.5.
Numerical results show clearly that, for small values of the
particle number, when gravity is negligible, the ratio mass to
particle number equals to $\omega$. For larger values of the
particle number gravity becomes more important and the negative
potential energy contributed by the attractive gravitational force
decreases the ratio $M$ to $N$. From fig. \ref{figure6} we also see
that the above ratio is larger for smaller values of the (negative)
cosmological constant. This reflects the repulsive character of a
negative cosmological constant. The star decreases its mass so as to
resist to the environment of the "negative" gravity induced by a
negative cosmological constant.

\vspace{1em}

\textbf{ACKNOWLEDGEMENTS}

\vspace{1em}

I wish to thank F. Brito for drawing my attention to the matter of
the entrapment of domain walls on the surface of soliton stars.


\begin{thebibliography}{06}
\bibitem{dw1}G. W. Gibbons and P. K. Townsend, Phys. Rev. Lett.
\textbf{83}, 1727 (1999).
\bibitem{dw2}P. M. Saffin, Phys. Rev. Lett. \textbf{83}, 4229
(1999).
\bibitem{dw3}D. Bazeia and F. A. Brito, Phys. Rev. \textbf{D 61},
105019 (2000).
\bibitem{dw4}S. M. Carroll, S. Hellerman and M. Trodden, Phys.
Rev. \textbf{D 61}, 065001 (2000).
\bibitem{dw-brane1}E. Witten, Nucl. Phys. \textbf{B 507}, 658
(1997).
\bibitem{dw-brane2}G. Dvali and A. Vilenkin, Phys. Rev. \textbf{D
67}, 046002 (2003).
\bibitem{dw-spherical}D. Bazeia and F. A. Brito, Phys. Rev
\textbf{D 62}, 101701(R) (2000).
\bibitem{dw-stars}F. Brito and D. Bazeia, Phys. Rev. \textbf{D
64}, 065022 (2001).
\bibitem{large-soliton-stars1}R. Friedberg,
T. D. Lee, and Y. Pang, Phys. Rev. D \textbf{35}, 3640 (1987).
\bibitem{large-soliton-stars2}R. Friedberg, T. D. Lee, and Y. Pang, Phys. Rev. D
\textbf{35}, 3658 (1987).
\bibitem{large-soliton-stars3}R. Friedberg, T. D. Lee, and Y. Pang, Phys. Rev. D
\textbf{35}, 3678 (1987).
\bibitem{large-soliton-stars4}T. D. Lee, Comments Nucl. Part. Phys.
\textbf{17}, 225 (1987).
\bibitem{dw-solitons}P. Sutcliffe, Phys. Rev. \textbf{D 68} 085004
(2003).
\bibitem{qstars-global}B. W. Lynn, Nucl. Phys. \textbf{B 321}, 465 (1989).
\bibitem{qstars-charged}A. Prikas, Phys. Rev. \textbf{D 66}, 025023 (2002).
\bibitem{qstars-extradim}A. Prikas, Phys. Rev. \textbf{D 69}, 125008 (2004).
\bibitem{qballs-initial}S. Coleman, Nucl. Phys. \textbf{B 262}, 263 (1985).
\bibitem{qballs-baryogen}K. Enqvist and J. McDonald, Phys. Lett.
\textbf{B 425}, 309 (1998).
\end{thebibliography}
\end{document}